\documentclass{iopart}
\usepackage{pstricks,graphicx,amsbsy,bbm,iopams}
\newcommand{\ket}[1]{\vert #1 \rangle} \newcommand{\bra}[1]{\langle #1 \vert}

\begin{document}
\title[Improving information/disturbance and estimation/distortion 
trade-offs with$\ldots$]{Improving information/disturbance and 
estimation/distortion 
trade-offs with non universal protocols}
\author{Stefano Olivares and Matteo G. A. Paris}
\address{Dipartimento di Fisica dell'Universit\`a di Milano, Italia.}
\date{\today}
%%%%%%%%%%%%%%%%%%%%%%%%%%%%%%%%%%%%%%%%%%%%%%%%%
\begin{abstract}
We analyze in details a conditional measurement scheme based on
linear optical components, feed-forward loop and homodyne
detection. The scheme may be used to achieve two different tasks.
On the one hand it allows the extraction of information with
minimum disturbance about a set of coherent states. On the other
hand, it represents a nondemolitive measurement scheme for the
annihilation operator, {\em i.e.} an indirect measurement of the
$Q$-function.  We investigate the information/disturbance
trade-off for state inference and introduce the
estimation/distortion trade-off to assess estimation of the
$Q$-function. For coherent states chosen from a Gaussian set we
evaluate both information/disturbance and estimation/distortion
trade-offs and found that non universal protocols may be
optimized in order to achieve better performances than universal
ones. For Fock number states we prove that universal protocols do
not exist and evaluate the estimation/distortion trade-off for a
thermal distribution.
\end{abstract}
%\pacs{}
%\keywords{}
%%%%%%%%%%%%%%%%%%%%%%%%%%%%%%%%%%%%%%%%%%%%%%%%%%%%%%%%%%%%%%%%%%%%%%
\section{Introduction}\label{s:intro}
Any measurement performed on a quantum systems alters the state
of system itself. As a consequence, any scheme aimed to extract {\em
information} about the state of a system unavoidably produces a {\em
disturbance}. The same is true if we focus on a specific quantity rather
than the state as a whole: any scheme devised for the {\em estimation}
of an observable or generalized observable produces a {\em
distortion} of the probability distribution of the measured quantity. 
\par
The trade-off between the amounts of the extracted information about a
quantum state and the corresponding added disturbance, from now on the
{\em information/disturbance} trade-off, has received much attention 
\cite{sc91,du98,bg96,fu96,fu01,KB1}.
Besides fundamental interest this is motivated by practical applications
in quantum communication and quantum cryptography \cite{gs02,gr1,gr2}. The
information/disturbance trade-off crucially depends on the set of input states 
and may be quantified in terms of fidelities. For finite-dimensional systems,
inequalities on fidelities, which express the bounds on precision imposed 
by quantum mechanics, have been derived in several cases. These include 
a single copy of
an unknown pure state \cite{KB1}, many copies of identically 
prepared pure qubits \cite{KB2}, a single copy of a pure state 
generated by independent
phase-shifts \cite{MistaPRA72}, an unknown spin coherent state \cite{msqph}, 
and single copy of an unknown 
maximally entangled state \cite{MS06}. 
Optimal measurement schemes, which saturate the bounds, have been
also devised \cite{MGenoni,MistaPRA05,sbqph} and implemented \cite{sc06}.
\par
A relevant concept in quantum estimation is universality. A protocol is
said to be universal if the fidelities are independent on the input
state, at least within the class of states under investigation. In fact, 
for systems with infinite-dimensional Hilbert space an  
information/disturbance trade-off has been derived for an unknown
coherent state ({\em i.e.} for a set of coherent states with flat
distribution of the amplitude) assuming universality and Gaussian
operations \cite{AndPRL06}. An optimal measurement scheme saturating this bound has also
been proposed and realized \cite{AndPRL06}.  In addition, it has been shown how to
slightly improve the trade-off using non Gaussian operations
\cite{MistaPRA73}. 
\par
The quantum mechanical back-action in the measurement of a specific
observable has been extensively studied in the context of quantum
nondemolition  measurements (QND) \cite{yama}. In a QND scheme, an observable is
measured, without destroying the state carrying the information, with
the aim of keeping the distortion (back-action) in the conjugated
observable and thus preserving the value of the observable itself \cite{grn}. 
The corresponding {\em estimation/distortion} trade-off has
been mostly analyzed in terms of variances. 
\par
In this paper we report a detailed analysis of an optical scheme based
on linear optical components and homodyne detection that can be used to
achieve two different, though related, tasks. 
On the one hand it allows to implement non
universal estimation protocol for Gaussian sets of coherent states and to improve
the trade-off, {\em i.e.} the extraction of information with minimum
disturbance, in comparison with universal protocols.  On the other hand,
it represents a nondemolitive measurement scheme of a generalized
observable, the annihilation operator, suitable for a generic set of states,
{\em i.e.} an indirect measurement of the $Q$-function.  We assess
its QND performances in terms of fidelities \cite{xqnd}, which quantify 
how much the measured distribution resembles the $Q$-function of the input 
states, and how much the distribution of the output states has been 
distorted by the measurement protocol. Going beyond variances allows us 
to investigate non Gaussian states. Indeed, we analyze the 
{\em estimation/distortion} trade-off for Gaussian sets of coherent states 
and thermal sets of Fock number states. 
\par
The scheme under investigation is that used in Ref. \cite{AndPRL06} 
to investigate the
universal Gaussian information/disturbance trade-off for an unknown 
coherent state. The same scheme has also been used to demonstrate 
$1\to 2$ optimal Gaussian cloning of coherent states \cite{AndPRL05} 
and suggested for more general cloning task, such as $1 \to m$ 
cloning of coherent states \cite{UlBook,Acta06} 
or cloning of general Gaussian states \cite{OPRA06}.
\par
The paper is structured as follows: In Section \ref{s:scheme} we
describe the measurement scheme as well its statistics and dynamics.
In Section \ref{s:est} we introduce the inference rules and the 
fidelities, whereas in Sections \ref{s:coh} and \ref{s:fock} we explicitly
evaluate the trade-offs for Gaussian sets of coherent states and thermal sets
of Fock number states, respectively. Section \ref{s:outro} closes the paper
with some concluding remarks.
%%%%%%%%%%%%%%%%%%%%%%%%%%%%%%%%%%%%%%%%%%%%%%%%%%%%%%%%%%%%%%%
\section{The measurement scheme}\label{s:scheme}
\begin{figure}[tb]
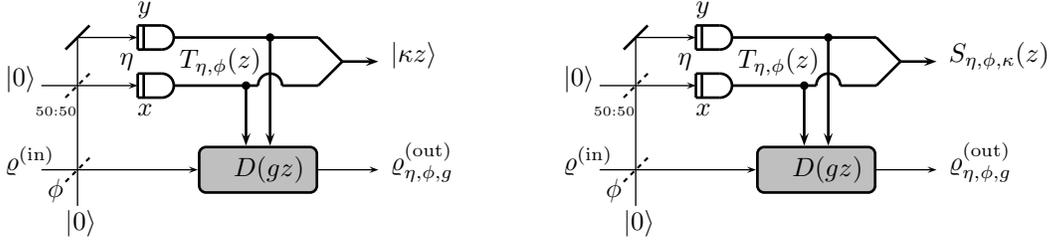

\begin{tabular}{ll}
\pspicture(0,0)(7,4)
%\psgrid(0,0)(0,0)(4,3)
\psset{unit=8mm}
\psline[linewidth=0.5pt]{->}(0.4,1)(3,1)
\psline[linewidth=0.5pt](1,0.4)(1,3.2)
\psline[linestyle=dashed,dash=2pt 2pt](0.8,0.8)(1.2,1.2)
\psline[linestyle=dashed,dash=2pt 2pt](0.8,2.2)(1.2,2.6)
\psline[linewidth=1pt](0.8,3)(1.2,3.4)
\psline[linewidth=0.5pt]{->}(0.4,2.4)(2,2.4)
\psline[linewidth=0.5pt]{->}(1,3.2)(2,3.2)
\psline[linewidth=1pt](2,2.2)(2,2.6)
\psline[linewidth=1pt](2.1,2.2)(2.1,2.6)
\psline[linewidth=1pt](2,2.2)(2.4,2.2)
\psline[linewidth=1pt](2,2.6)(2.4,2.6)
\psarc[linewidth=1pt](2.4,2.4){0.2}{-90}{90}
\psline[linewidth=1pt](2,3)(2,3.4)
\psline[linewidth=1pt](2.1,3)(2.1,3.4)
\psline[linewidth=1pt](2,3)(2.4,3)
\psline[linewidth=1pt](2,3.4)(2.4,3.4)
\psarc[linewidth=1pt](2.4,3.2){0.2}{-90}{90}
\psline[linewidth=1pt](2.6,2.4)(4,2.4)
\psline[linewidth=1pt](4.4,2.4)(5,2.4)
\psline[linewidth=1pt]{->}(3.8,2.4)(3.8,1.4)
\psarc[linewidth=1pt](4.2,2.4){0.2}{0}{180}
\pscircle*(3.8,2.4){0.07}
\psline[linewidth=1pt](2.6,3.2)(5,3.2)
\psline[linewidth=1pt]{->}(4.2,3.2)(4.2,1.4)
\pscircle*(4.2,3.2){0.07}
\psframe[framearc=0.3,linewidth=1pt,fillstyle=solid,fillcolor=lightgray](3,0.6)(5,1.4)
\psline[linewidth=1pt](5,3.2)(5.4,2.8)
\psline[linewidth=1pt](5,2.4)(5.4,2.8)
\psline[linewidth=1pt]{->}(5.4,2.8)(6,2.8)
\psline[linewidth=0.5pt]{->}(5,1)(6,1)
\put(0.5,0.55){$\phi$}
\put(0.3,1.9){\tiny 50:50}
\put(-0.2,1){$\varrho^{\rm (in)}$}
\put(-0.2,2.4){$\ket{0}$}
\put(0.8,0){$\ket{0}$}
\put(3.6,0.9){$D(gz)$}
\put(6.2,1){$\varrho^{\rm (out)}_{\eta,\phi,g}$}
\put(6.2,2.8){$\ket{\kappa z}$}
\put(2,1.9){$x$}
\put(2,3.6){$y$}
\put(1.7,2.75){$\eta$}
\put(2.7,2.7){$T_{\eta,\phi}(z)$}
\endpspicture
&
\pspicture(0,0)(7,4)
\psset{unit=8mm}
%\psgrid(0,0)(0,0)(4,3)
\psline[linewidth=0.5pt]{->}(0.4,1)(3,1)
\psline[linewidth=0.5pt](1,0.4)(1,3.2)
\psline[linestyle=dashed,dash=2pt 2pt](0.8,0.8)(1.2,1.2)
\psline[linestyle=dashed,dash=2pt 2pt](0.8,2.2)(1.2,2.6)
\psline[linewidth=1pt](0.8,3)(1.2,3.4)
\psline[linewidth=0.5pt]{->}(0.4,2.4)(2,2.4)
\psline[linewidth=0.5pt]{->}(1,3.2)(2,3.2)
\psline[linewidth=1pt](2,2.2)(2,2.6)
\psline[linewidth=1pt](2.1,2.2)(2.1,2.6)
\psline[linewidth=1pt](2,2.2)(2.4,2.2)
\psline[linewidth=1pt](2,2.6)(2.4,2.6)
\psarc[linewidth=1pt](2.4,2.4){0.2}{-90}{90}
\psline[linewidth=1pt](2,3)(2,3.4)
\psline[linewidth=1pt](2.1,3)(2.1,3.4)
\psline[linewidth=1pt](2,3)(2.4,3)
\psline[linewidth=1pt](2,3.4)(2.4,3.4)
\psarc[linewidth=1pt](2.4,3.2){0.2}{-90}{90}
\psline[linewidth=1pt](2.6,2.4)(4,2.4)
\psline[linewidth=1pt](4.4,2.4)(5,2.4)
\psline[linewidth=1pt]{->}(3.8,2.4)(3.8,1.4)
\psarc[linewidth=1pt](4.2,2.4){0.2}{0}{180}
\pscircle*(3.8,2.4){0.07}
\psline[linewidth=1pt](2.6,3.2)(5,3.2)
\psline[linewidth=1pt]{->}(4.2,3.2)(4.2,1.4)
\pscircle*(4.2,3.2){0.07}
\psframe[framearc=0.3,linewidth=1pt,fillstyle=solid,fillcolor=lightgray](3,0.6)(5,1.4)
\psline[linewidth=1pt](5,3.2)(5.4,2.8)
\psline[linewidth=1pt](5,2.4)(5.4,2.8)
\psline[linewidth=1pt]{->}(5.4,2.8)(6,2.8)
\psline[linewidth=0.5pt]{->}(5,1)(6,1)
\put(0.5,0.55){$\phi$}
\put(0.3,1.9){\tiny 50:50}
\put(-0.2,1){$\varrho^{\rm (in)}$}
\put(-0.2,2.4){$\ket{0}$}
\put(0.8,0){$\ket{0}$}
\put(3.6,0.9){$D(gz)$}
\put(6.2,1){$\varrho^{\rm (out)}_{\eta,\phi,g}$}
\put(6.2,2.8){$S_{\eta,\phi,\kappa}(z)$}
\put(2,1.9){$x$}
\put(2,3.6){$y$}
\put(1.7,2.75){$\eta$}
\put(2.7,2.7){$T_{\eta,\phi}(z)$}
\endpspicture
\end{tabular}
\caption{\label{f:scheme} Linear optical schemes for state inference (left)
and indirect measurement of the $Q$-function (right). 
In both schemes the input signal $\varrho^{\rm (in)}$ impinges onto a 
beam splitter with transmissivity $\tau = \cos^2\phi$: the reflected 
part is measured by double homodyne detection, and the complex measurement 
outcome $z = x + iy$ is used to displace the transmitted beam by an amount 
$gz$, $g$ being a suitable gain factor.
In the state inference scheme (left), the measurement outcome is used to
infer the input state according to the rule $z \leadsto \ket{\kappa z}$,
$\kappa$ being a real number and $|z\rangle$ a coherent state. In 
the estimation scheme (right) the measurement outcomes are collected to 
form the distribution $S_{\eta,\phi,\kappa}(z)$, which is used as an estimate
of the $Q$-function of the input signal. See text for more details.}\end{figure}

The measurement scheme we are going to analyze is schematically depicted 
in Fig.~\ref{f:scheme}, where we show the configuration used for state inference 
(left) as well as that used for nondemolitive measurement of the $Q$-function 
(right). In our scheme, the signal $\varrho^{\rm (in)}$, {\em i.e.} mode $1$, is mixed
with the vacuum at a beam splitter with transmissivity $\tau = \cos^2 \phi$. 
The reflected beam is then measured by a double homodyne detector with quantum 
efficiency $\eta$. The outcomes of the measurement are complex numbers
$z=x+iy$, $x$ and $y$ being the outcomes from the two homodyne detectors,
which are used either to infer the state at the input, or collected
to build an estimate of the input $Q$-function. In both cases, a suitable
real rescaling factor $\kappa$ may be used to optimize the fidelities.  
The outcome of the measurement is also sent to the transmitted beam, which 
is displaced by an amount $gz$, $g$ being an suitable additional gain.
The positive operator-valued measure (POVM) of the double homodyne is 
given by 
\begin{equation}
\Pi_\eta(z) = \int_{\mathbb C} \frac{d^2\mu}{\pi \Delta_\eta^2}\,,
\exp\left\{-\frac{|\mu - z|^2}{\Delta_\eta^2}\right\}\,
\frac{\ket{\mu}\bra{\mu}}{\pi}\,,
\end{equation}
with $\Delta_\eta^2 = (1-\eta)/\eta$, $\eta$ being the quantum 
efficiency of each detector (here we assume that both the detectors have
the same efficiency).  The probability distribution of the raw outcomes $z$
is given by
\begin{equation}
T_{\eta,\phi}(z) = {\rm Tr}_{12}\left[
U_\phi\,\varrho^{\rm (in)}\otimes\ket{0}\bra{0}\,U_\phi^{\dag}\,
{\mathbbm 1}\otimes \Pi_\eta(z) \right]\,,
\end{equation}
where $U_\phi=\exp\{\phi(a_1^\dag a_2 - a_1 a^\dag_2)\}$ is the evolution 
operator of the beam splitter. The conditional state of mode $1$, after the 
outcome $z$ is given by
\begin{equation}
\varrho_{\eta,\phi}(z) = \frac{
{\rm Tr}_{2}\left[U_\phi\,\varrho^{\rm (in)}\otimes\ket{0}\bra{0}\,U_\phi^{\dag}\,
{\mathbbm 1}\otimes \Pi_\eta(z) \right]
}{T_{\eta,\phi}(z)}\,,
\end{equation}
and the overall output state $\varrho^{\rm (out)}_{\eta,\phi,g}$
is obtained averaging over all the possible outcomes
\begin{equation}
\varrho^{\rm (out)}_{\eta,\phi,g} = \int_{\mathbb C}
d^2z\,T_{\eta,\phi}(z)\,D(gz)\,\varrho_{\eta,\phi}(z)\,D^{\dag}(gz)\,,
\end{equation}
with $D(gz)=\exp\{gza^\dag - g z^* a\}$.
\par
Using the Glauber-Sudarshan $P$-function representation, we can write 
the input state as 
\begin{equation}
\varrho^{\rm (in)} = \int_{\mathbb C} d^2\xi\, P^{\rm (in)}(\xi)\,
\ket{\xi}\bra{\xi}\,.
\end{equation}
In turn, the probability distribution of the outcomes may be written as 
\begin{equation}\label{T:wig:s}
T_{\eta,\phi}(z) = \frac{1}{\sin^2\phi}\, W_{s_1}[\varrho^{\rm (in)}]
\left( \frac{z}{\sin\phi}
\right)\,,
\end{equation}
with
\begin{equation}
s_1 = 1 - \frac{2}{\eta \sin^2\phi}\,,
\end{equation}
where $W_s[\varrho](\alpha)$ denotes the $s$-ordered Wigner function
of the state $\varrho$ \cite{cah}. We also made use of the relation, valid for $r>s$, 
\begin{equation}\label{w:r-s}
W_s[\varrho](\zeta) = \int_{\mathbb C} d^2\xi\, \frac{2}{\pi
(r-s)}\,\exp\left\{-\frac{2|\xi -\zeta|^2}{r-s}\right\}\, W_r[\varrho](\xi)\,.
\end{equation}
%%%%%%%%%%%%%%
%%%%%%%%%%%%%%
In addition, it is straightforward to prove that 
\begin{equation}\label{rho:out}
\varrho^{\rm (out)}_{\eta,\phi,g} = \int_{\mathbb C} d^2\xi\,
P^{\rm (out)}_{\eta,\phi,g}(\xi)\, \ket{\xi}\bra{\xi}\,, 
\end{equation}
with
\begin{equation}
P^{\rm (out)}_{\eta,\phi,g}(\xi) = \frac{1}{(\cos\phi + g \sin\phi)^2}\,
W_{s_2}[\varrho^{\rm (in)}]\left( \frac{\xi}{\cos\phi + g \sin\phi} \right)\,,
\end{equation}
and
\begin{equation}
s_2 = 1 - \frac{2g^2}{\eta(\cos\phi + g\sin\phi)^2}\,.
\end{equation}
Finally, thanks to Eq.~(\ref{w:r-s}), the $Q$-function at the output, 
corresponding to the state (\ref{rho:out}), can be written as
\begin{equation}
Q^{\rm (out)}_{\eta,\phi,g}(z) = \frac{1}{(\cos\phi + g\sin\phi)^2}
W_{s_3}[\varrho^{\rm (in)}]\left( \frac{1}{\cos\phi + g\sin\phi}  \right)\,,
\end{equation}
with
\begin{equation}
s_3 = 1 - \frac{2(\eta - g^2)}{\eta(\cos\phi + g\sin\phi)^2}\,.
\end{equation}
%%%%%%%%%%%%%%%%%
\section{Inferences and fidelities}\label{s:est}
In this Section we introduce inference rules and fidelities to quantify 
the information/disturbance trade-off for state inference and the
estimation/distortion trade-off for measurement of the $Q$-function.
\subsection{State inference: information fidelity and disturbance fidelity}
If the input signal belongs to a Gaussian set of coherent states 
then the reflected beam is still a coherent state and a natural
inference rule \cite{Helstrom} after having observed the outcome $z$ 
is the following: $z\rightsquigarrow \ket{\kappa z}$, with $\kappa \ge 0$.
In order to assess our inference, assuming a set of pure states at the
input, we use the state overlap between the inferred state and the input
one.  By averaging over the possible outcomes $z$, we arrive at the {\em
information fidelity}:
\begin{equation}\label{information}
G_{\eta,\kappa}(\phi) =
\int_{\mathbb C} d^2z\, T_{\eta,\phi}(z)\,
\bra{\kappa z}\varrho^{\rm (in)}\ket{\kappa z}\,.
\end{equation}
which can be optimized, i.e., maximized, with respect to the parameter $\kappa$.
Similarly, the amount of disturbance can be evaluated by the overlap between 
the input state  and the conditional one. By averaging over the possible
outcomes $z$ we have the {\em disturbance fidelity}:
\begin{equation}
\label{disturbance}
\!\!\!\!\!\!\!\!\!\!\!
\!\!\!\!\!\!\!\!\!\!\!
F_{\eta,g}(\phi) =
\int_{\mathbb C} d^2z\, T_{\eta,\phi}(z)\,
{\rm Tr}\left[\varrho^{\rm (in)}\, 
D(gz)\:\varrho_{\eta,\phi}(z)\:D^\dag (z) \right]=
{\rm Tr}\left[\varrho^{\rm (in)}\, 
\varrho^{\rm (out)}_{\eta,\phi,g}\right]\,,
\end{equation}
where, again, we assumed pure states at the input.
The disturbance fidelity can be optimized, i.e., maximized, 
with respect to the parameter $g$.
%%%%%%
\subsection{Measurement of the $Q$-function: estimation fidelity and distortion fidelity}
Performing double-homodyne detection on the reflected beam provides an estimate of the
input $Q$-function upon a suitable rescaling of raw outcomes. As an estimate 
we adopt the distribution $S_{\eta,\phi,\kappa}(z)$, defined as follows 
\begin{equation}
S_{\eta,\phi,\kappa}(z) = \frac{1}{\kappa^2}\,
T_{\eta,\phi}\left( \frac{z}{\kappa} \right)\,,
\end{equation}
with $\kappa \ge 0$ (see right panel of Fig.~\ref{f:scheme}). In order to evaluate the
similarity of the inferred $Q$-function to the input one
$Q^{\rm (in)}(z) = \frac{1}{\pi}\bra{z} \varrho^{\rm (in)} \ket{z}$ 
we introduce the {\em estimation fidelity}:
\begin{equation}\label{estiamtion}
H_{\eta,\kappa}(\phi) =
\int_{\mathbb C} d^2z\, \sqrt{Q^{\rm (in)}(z)\,S_{\eta,\phi,\kappa}(z)}\,.
\end{equation}
$H_{\eta,\kappa}(\phi)$ is a proper fidelity, {\em i.e.}
$0\leq H_{\eta,\kappa}(\phi) \leq 1$, with $H_{\eta,\kappa}(\phi)=1$ iff
the inferred distribution is equal to the actual $Q$-function.
The protocol can be optimized, by maximizing
$H_{\eta,\kappa}(\phi)$ with respect to the parameter $\kappa$.  Since the
output state is altered by the measurement, the corresponding
$Q$-function, $Q^{\rm (out)}_{\eta,\phi,g}(z)$, is a distorted version of
input one. The degree of this modification can be evaluated by means of the
{\em distortion fidelity}
\begin{equation}\label{distortion}
K_{\eta,g}(\phi) =
\int_{\mathbb C} d^2z\, \sqrt{Q^{\rm (in)}(z)\,
Q_{\eta,\phi,g}^{\rm (out)}(z)}\,,
\end{equation}
which can be optimized, {\em i.e.}, maximized, with respect to the parameter $g$.
%%%%%
\section{Coherent states}\label{s:coh}
In this Section we evaluate explicitly the information/disturbance 
and the estimation/distortion trade-offs for a set
of coherent states $\varrho^{\rm (in)} = \ket{\beta}\bra{\beta}$
with complex amplitudes distributed according to the Gaussian
\begin{equation}
{\cal P}(\beta) = \frac{1}{\pi \Omega^2}\,
\exp\left\{-\frac{|\beta|^2}{\Omega^2}\right\}\:.
\end{equation}
Such a distribution of coherent states can be obtained, {e.g.}, starting
from the single output states of a continuous variable teleportation protocol
as well as at the output of a Gaussian noise channel with vacuum input.
%%%
\subsection{Information/disturbance trade-off}
Since the Glauber $P$-function of a coherent state 
$P^{\rm (in)}(\xi) = \delta^{(2)}(\xi-\beta)$ is a 
delta function in the complex plane 
the information fidelity is given by 
\begin{equation}
G_{\eta,\kappa}(\phi,\beta) = \frac{\eta}{\eta + \kappa^2}\,
\exp\left\{ - \frac{\eta (1-\kappa\sin\phi)^2}{\eta + \kappa^2}|\beta|^2 \right\}\,,
\end{equation}
The protocol is universal ({\em i.e.} $G$ does not depend on $\beta$) 
if $\kappa = 1/\sin\phi$; the corresponding fidelity is given by
\begin{equation}
G_{\eta}(\phi) = \frac{\eta\sin^2\phi}{1 + \eta\sin^2\phi}\,.
\end{equation}
For the disturbance fidelity we have
\begin{eqnarray}
F_{\eta,g}(\phi,\beta) = \frac{\eta}{\eta + g^2} 
\exp\left\{ - \frac{\eta (1 - \cos\phi - g \sin\phi)^2}{\eta +
g^2}|\beta|^2 \right\}\,.
\end{eqnarray}
Universality is obtained for $g = (1-\cos\phi)/\sin\phi$, {\em i.e.},
\begin{equation}
F_{\eta}(\phi) = \frac{\eta \sin^2\phi}
{\eta\sin^2\phi + (1-\cos\phi)^2}\,.
\end{equation}
The (universal) information/disturbance trade-off reads as follows
\begin{equation}
F_\eta = G \left\{
G + (1-G)\left[
1-\sqrt{1-\frac{G}{\eta(1-G)}}
\right]^2
\right\}^{-1}
%\frac{G}{2(1-G-\sqrt{(1-G)[\eta(1-G)-G]/\eta})}
\label{FGuniv}\;.
\end{equation}
For $\eta \rightarrow 1$ we recover the optimal trade-off obtained in Ref.
\cite{AndPRL06}. Notice that also for $\eta\neq 1$ the trade-off
(\ref{FGuniv}) is optimal, {\em i.e.}, the noise added is the minimum
allowed by quantum mechanics in a joint measurement of conjugated
quadratures  \cite{OPRA06}. As expected we have $F_\eta < F_1$, $\forall
G,\eta$, {\em i.e.}, a non unit value of the quantum efficiency degrades
performances. 
\par
In the general case, {\em i.e.}, releasing the request of universality,
the average information and disturbance fidelities are given by
\begin{eqnarray}
\!\!\!\!\!\!\!\!\!\!\!
\!\!\!\!\!\!\!\!\!\!\!
\overline{G}_{\eta,\kappa}(\phi) = \int_{\mathbb C} d^2\beta\,
{\cal P}(\beta)\, G_{\eta,\kappa}(\phi,\beta) = \frac{\eta}
{\eta + \kappa^2 + \eta \Omega^2 (1-\kappa\sin\phi)^2}\,,
\\
\!\!\!\!\!\!\!\!\!\!\!
\!\!\!\!\!\!\!\!\!\!\!
\overline{F}_{\eta,g}(\phi) = \int_{\mathbb C} d^2\beta\,
{\cal P}(\beta)\, F_{\eta,g}(\phi,\beta)  = \frac{\eta}
{\eta + g^2 + \eta \Omega^2 (1- \cos\phi - g\sin\phi)^2}\,,
\end{eqnarray}
respectively.
The fidelity may be maximized with respect to the parameters
$\kappa$ and $g$, whose optimal values are given by
\begin{equation}
\kappa = \frac{\eta \Omega^2 \sin\phi}{1 + \eta \Omega^2 \sin^2\phi}
\qquad g = \frac{\eta \Omega^2 \sin^2\phi\,(1-\cos\phi)}
{1 + \eta \Omega^2 \sin^2\phi}\,,
\end{equation}
corresponding to 
\begin{eqnarray}
\overline{G}_{\eta,\Omega}(\phi) &=
\frac{1 + \eta \Omega^2 \sin^2\phi}
{1 + \Omega^2 + \eta \Omega^2 \sin^2\phi}\,.
\nonumber \\
\overline{F}_{\eta,\Omega}(\phi) &=
\frac{1 + \eta \Omega^2 \sin^2\phi}
{1 + \Omega^2 \left[ \eta\sin^2\phi
+\left( 1 - \cos\phi \right)^2 \right]}\,.
\end{eqnarray}
and to the trade-off 
\begin{equation}
\!\!\!\!\!\!\!\!\!\!\!
\!\!\!\!\!\!\!\!\!\!\!
\overline{F}_{\eta,\Omega} =
\frac{(1+\Omega^2)\overline{G}}{\Omega^2}
\left\{
\overline{G}+(1-\overline{G})
\left[
1-\sqrt{1-\frac{(1+\Omega^2)\overline{G}-1}{\eta\Omega^2(1-\overline{G})}}
\right]^2
\right\}^{-1}
%\frac{\overline{G}}{\overline{G}
%%+(1-\overline{G})\left(1-\sqrt{1-\frac{\overline{G}}{\eta \Omega^2 
%(1-\overline{G})}}\right)^2}
\label{FGnu}\;.
\end{equation}
For ``large'' set of signals, {\em i.e.}, for $\Omega \rightarrow \infty$, we
recover the ``universal'' trade-off (\ref{FGuniv}), whereas for finite
values of $\Omega$ we have $\overline{F}_{\eta,\Omega} > F_{\eta}$. In
other words, for finite $\Omega$ non universal protocols may be optimized
and achieve superior performances compared to universal one.
\begin{figure}[h!]
\includegraphics[width=0.45\textwidth]{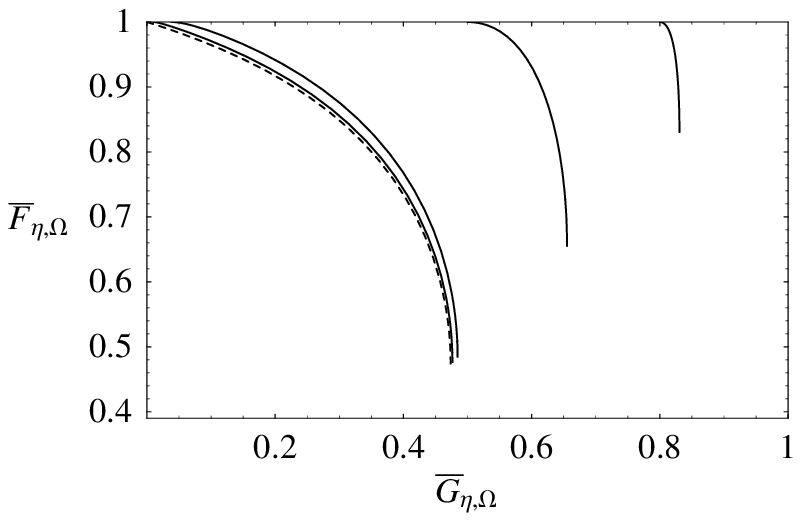}
\includegraphics[width=0.45\textwidth]{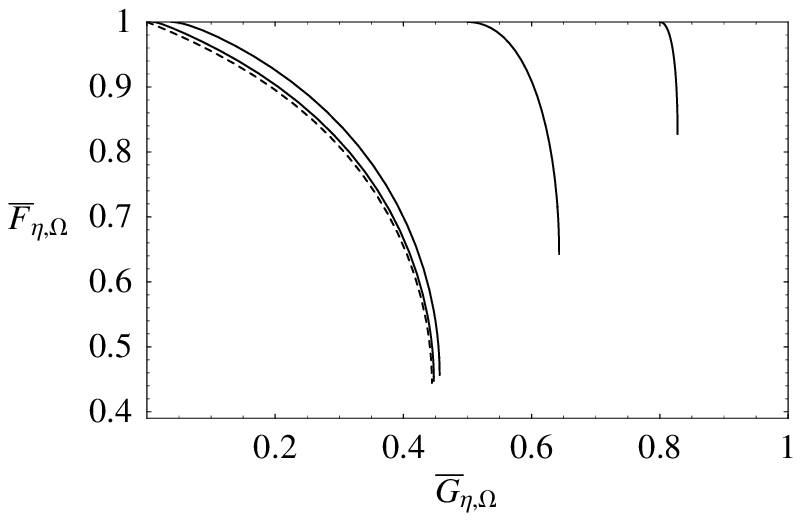}
\caption{\label{f:GF} Information/disturbance trade-off for universal and
non universal protocols. We show the disturbance fidelity as a function 
of the information fidelity for different values of the width $\Omega$ and 
for two different values of the quantum efficiency (see text for details).
(Left): $\eta=0.9$. (Right): $\eta=0.8$. The dashed line refers to universal
protocol whereas, in both the plots, the solid lines are for (from right to
left): $\Omega=0.5,1.0,5.0$ and $10.0$.}
\end{figure} \par\noindent
In Fig.~\ref{f:GF} we plot the information/disturbance trade-off, which is
obtained by tuning $\phi$ in the interval $[0,\pi/2]$ and, in turn, the
transmissivity $\tau=\cos^2\phi$ of the beam splitter ranges from $1$ to
$0$.  When $\tau=1$, the input states are completely transmitted and 
only the vacuum is left for the double homodyne detection: in this
case the disturbance fidelity $\overline{F}_{\eta,\Omega} = 1$ while
the information fidelity $\overline{G}_{\eta,\Omega}$ reaches its
minimum. When $\tau=0$, the input state is completely reflected and 
nothing is transmitted: now $\overline{F}_{\eta,\Omega}$ is minimum and
$\overline{G}_{\eta,\Omega}$ reaches its maximum.
The universal protocol has been recently experimentally
demonstrated in \cite{AndPRL06}, were the quantum efficiency was 
approximately $\eta\approx 94\%$.
%%%
\subsection{Estimation/distortion trade-off}
\begin{figure}[h]
\includegraphics[width=0.45\textwidth]{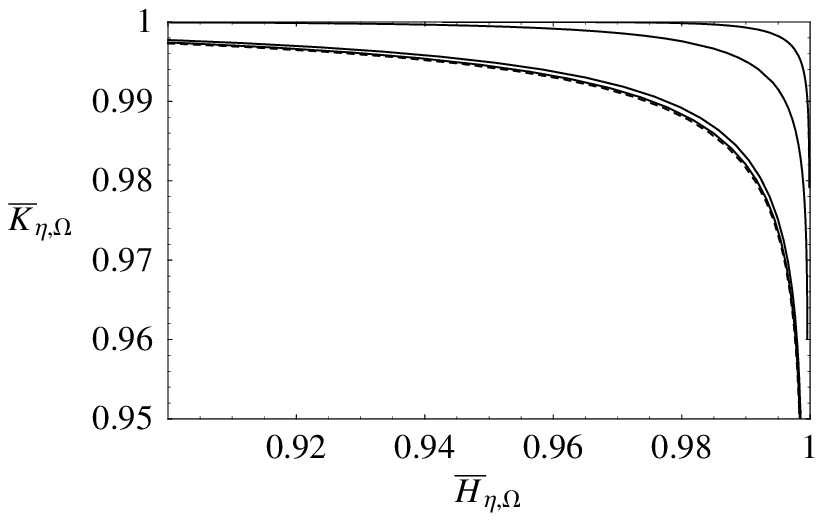}
\includegraphics[width=0.45\textwidth]{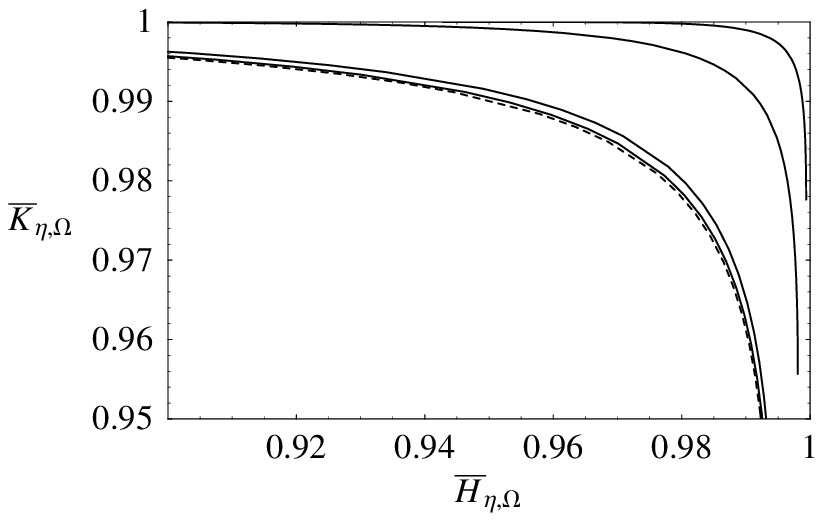}
\caption{\label{f:HK} Average estimation/distortion trade-off for
different values of $\Omega$ and for two different values of the quantum
efficiency (see text for details). (Left): $\eta=0.9$. (Right): $\eta=0.8$. 
The dashed line is the universal trade-off whereas, in both the plots, 
the solid lines are for (from right to left): $\Omega=0.5,1.0,5.0$ and $10.0$.}
\end{figure}

The estimation and distortion fidelities for a given coherent state read
as follows
\begin{eqnarray}
H_{\eta,\kappa}(\phi,\beta) &= 
\frac{2 \kappa \sqrt{\eta}}{\eta + \kappa^2}\,
\exp\left\{
-\frac{\eta (1 - \kappa \sin\phi)^2}{2(\eta + \kappa^2)}|\beta|^2
\right\}\\
K_{\eta,g}(\phi,\beta) &= \frac{2\sqrt{\eta(\eta + g^2)}}{2\eta + g^2}
\exp\left\{
\frac{\eta(1 - \cos\phi - g\sin\phi)^2}{2(2\eta + g^2)}|\beta|^2
\right\}\,,
\end{eqnarray}
Universality conditions are given by  $\kappa = 1/\sin\phi$ and $g =
(1-\cos\phi)/\sin\phi$, and The corresponding universal fidelities are 
\begin{eqnarray}
H_{\eta}(\phi) &= \frac{2\sqrt{\eta}\,
\sin\phi}{1+\eta\sin^2\phi} \\ 
K_{\eta}(\phi) &= \frac{2 \sin\phi\, \sqrt{\eta\left[\eta \sin^2\phi
+(1-\cos\phi)^2\right]}}{2\eta\sin^2\phi + (1-\cos\phi)^2}\,.
\end{eqnarray}
For non universal protocols we have
\begin{eqnarray}
\!\!\!\!\!\!\!\!
\!\!\!\!\!\!\!\!
\!\!\!\!\!\!\!\!
\!\!\!\!\!\!\!\!
\overline{H}_{\eta,\kappa}(\phi) = \int_{\mathbb C} d^2\beta\,
{\cal P}(\beta)\, H_{\eta,\kappa}(\phi,\beta) = \frac{4 \kappa \sqrt{\eta}}
{2(\eta + \kappa^2) + \eta \Omega^2 (1+k\sin\phi)^2}\,,
\\
\!\!\!\!\!\!\!\!
\!\!\!\!\!\!\!\!
\!\!\!\!\!\!\!\!
\!\!\!\!\!\!\!\!
\overline{K}_{\eta,g}(\phi) = \int_{\mathbb C} d^2\beta\,
{\cal P}(\beta)\, K_{\eta,g}(\phi,\beta) = 
\frac{4\sqrt{\eta(\eta + g^2)}}
{2(2\eta + g^2) + \eta \Omega^2 (1- \cos\phi - g\sin\phi)^2}\,.
\end{eqnarray}
The estimation fidelity is maximized for
\begin{equation}
\kappa = \sqrt{\frac{\eta(2+\Omega^2)}
{2 + \eta \Omega^2 \sin^2\phi}}\,,
\end{equation}
whereas the distortion fidelity is maximized when $g$ is equal
to the real root of the following cubic equation
\begin{equation}
\fl g^3 \left( 2 + \eta \Omega^2 \sin^2\phi \right)
  +g \eta  \Omega ^2 \left[2 \eta  \sin ^2(\phi )-(1-\cos\phi)^2 \right] =
  \eta ^2 \Omega ^2 (2 \sin \phi - \sin2\phi)\,.
\end{equation}
The optimized estimation fidelity is given by
\begin{equation}
\fl \overline{H}_{\eta} (\phi) =
\frac{2\sqrt{\left( 2+\Omega^2\right)\left( 2 + \eta \Omega^2 \sin^2\phi \right) }}
{\left( 2+\Omega^2\right)\left( 2 + \eta \Omega^2 \sin^2\phi \right)
-\Omega^2 \sin\phi
\sqrt{\eta\left( 2+\Omega^2\right)\left( 2 + \eta \Omega^2 \sin^2\phi
\right)}}\,,
\end{equation}
whereas we do not report the analytic expression for the optimized 
distortion fidelity $\overline{K}_{\eta}(\phi)$ which is quite cumbersome.
In Fig.~\ref{f:HK} we show the estimation/distortion trade-off 
for different values of the quantum efficiency $\eta$ and the 
width $\Omega$ of the distribution. 
The universal trade-off is recovered for $\Omega\rightarrow \infty$.
\section{Fock number states}\label{s:fock}
\begin{figure}[tb]
\begin{tabular}{ll}
\pspicture(0.1,0)(7,4)
%\psgrid(0,0)(0,0)(4,3)
\psset{unit=1mm}
\put(0,0){\includegraphics[width=0.45\textwidth]{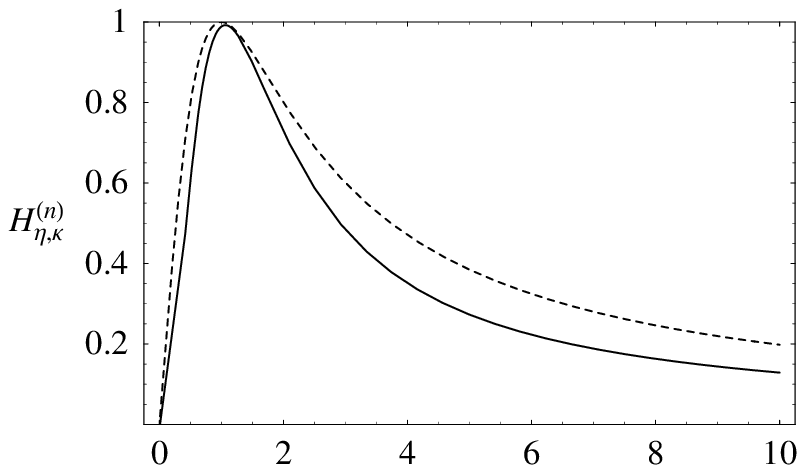}}
\put(34,20){\includegraphics[width=0.16\textwidth]{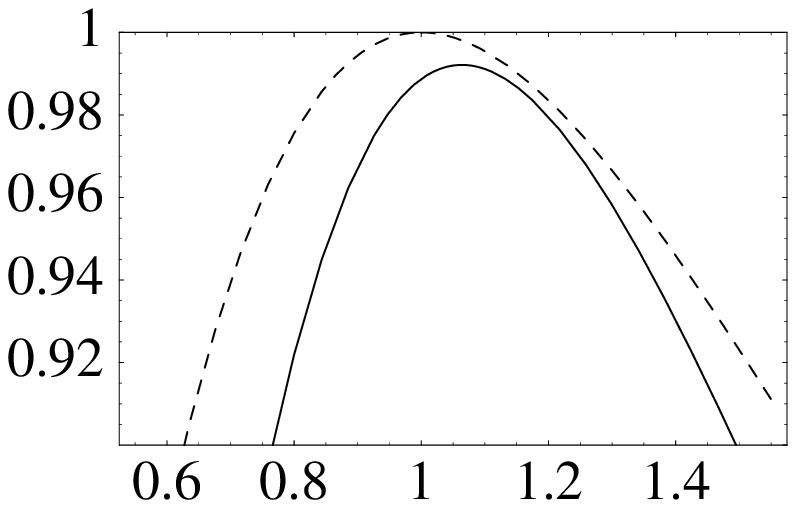}}
\put(33,2){\small $\kappa$}
\endpspicture
&
\pspicture(0.9,0)(7,4)
\psset{unit=1mm}
%\psgrid(0,0)(0,0)(4,3)
\put(0,0){\includegraphics[width=0.45\textwidth]{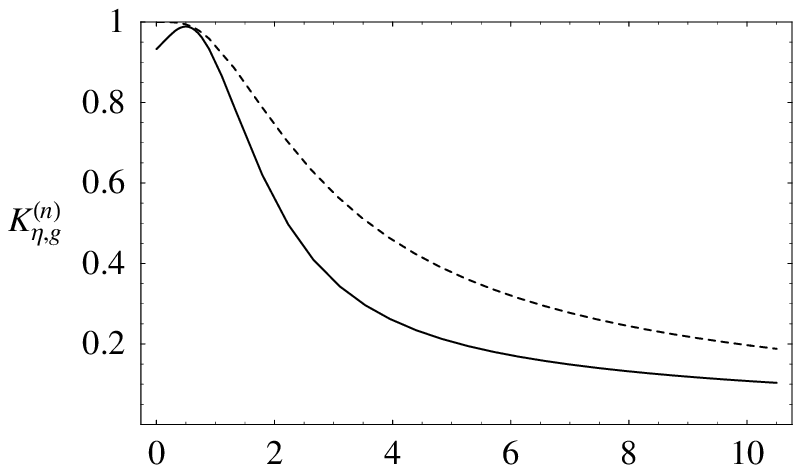}}
\put(34.5,20){\includegraphics[width=0.16\textwidth]{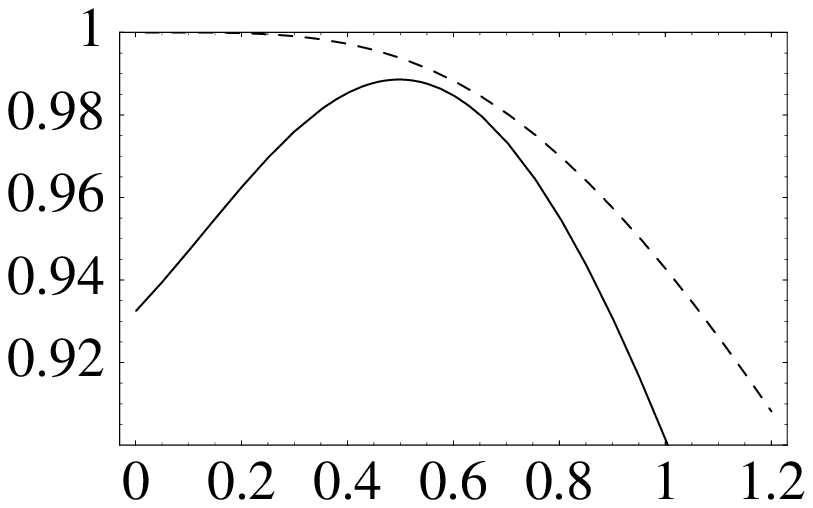}}
\put(33,2){\small $g$}
\endpspicture
\end{tabular}
\vspace{-0.3cm}
\caption{\label{f:NonUni}
(Right) Plots of $H_{\eta,\kappa}^{(n)}(\phi)$ as a function of the
parameter $\kappa$; (Left) plots of $K_{\eta,g}^{(n)}(\phi)$ as a function
of the parameter $g$. In both the plots we set $\phi=\pi/3$, $\eta=1$ and we have
chosen $n=0$ (dashed lines) and $n=1$ (solid lines). The insets are
magnification of the regions nearby the maxima. Notice that there are not
intersections.}
\end{figure}

\begin{figure}[h]
\includegraphics[width=0.45\textwidth]{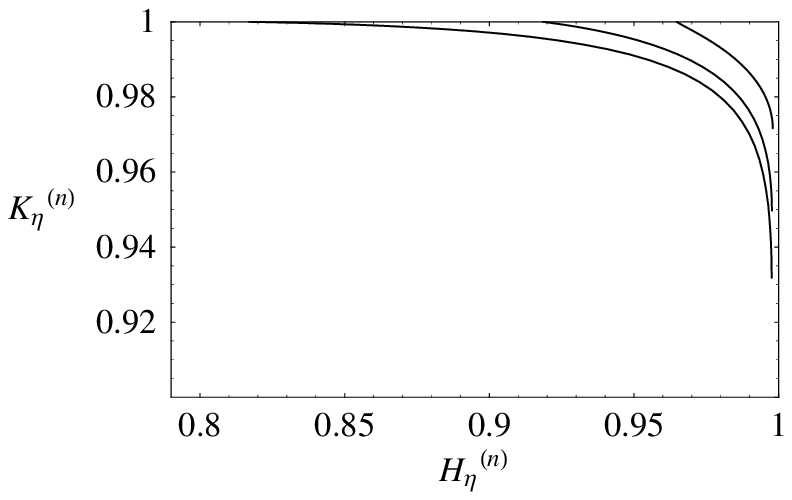}
\includegraphics[width=0.45\textwidth]{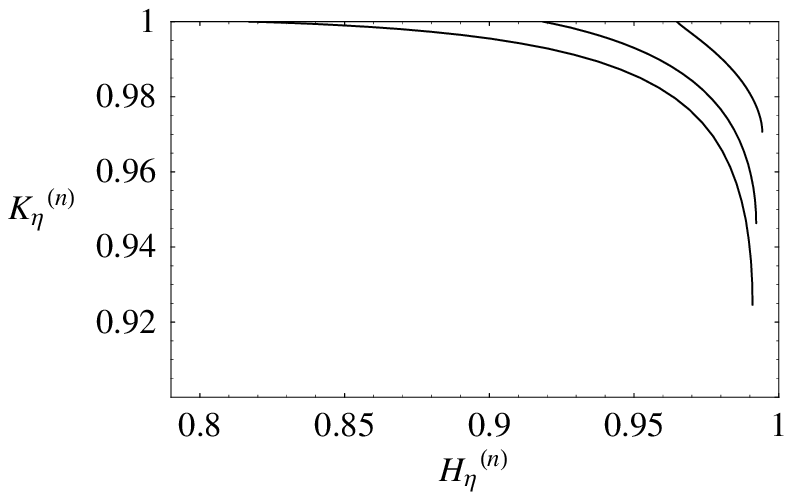}
\caption{Optimized estimation/distortion trade-off for 
Fock number states. (Left): $\eta=0.9$. (Right): $\eta=0.8$.
In both plots we show the trade-off for (from right to left)
$n=1,2,5$.\label{f:n}}
\end{figure}

Since the $s$-ordered Wigner function of the Fock state $|n\rangle$ is
given by
\begin{eqnarray}
\!\!\!\!\!\!\!\!
\!\!\!\!\!\!\!\!
W^{(n)}_s(\xi) =
(-1)^n \frac{2}{\pi(1-s)}\left(\frac{1+s}{1-s}\right)^n
\exp\left\{-\frac{2|\xi|^2}{1-s} \right\}\: 
L_n\left( \frac{4|\xi|^2}{1-s^2} \right)\,,
\end{eqnarray}
where $L_n(z)$ are Laguerre polynomials, we can evaluate the estimation and
distortion fidelities $H_{\eta,\kappa}^{(n)}(\phi)$, $K_{\eta,g}^{(n)}(\phi)$
respectively, as described in Sections \ref{s:scheme} and \ref{s:est}.
For Fock states universal protocols, {\em i.e.}, protocols independent of 
$n$ do not exist. To prove this let us consider the simple case of 
$\ket{0}$ and $\ket{1}$ as input states. Universality would require 
that $ \forall \eta,\phi$ there exist $\overline{\kappa} = 
\overline{\kappa}(\eta,\phi) $ and $\overline{g} =
\overline{g}(\eta,\phi)$ such that
$H_{\eta,\overline{\kappa}}^{(0)}(\phi)=
H_{\eta,\overline{\kappa}}^{(1)}(\phi)$ and $
K_{\eta,\overline{g}}^{(0)}(\phi)=
K_{\eta,\overline{g}}^{(1)}(\phi)$. 
On the other hand, if we set, for example, $\phi=\pi/3$ and $\eta=1$,
the two conditions are never satisfied, as shown in
Fig.~\ref{f:NonUni}. 
%%%%%%%%%%%%%%
%%%%%%%%%%%%%%
%%
In Fig. \ref{f:n} we show the optimized estimation/distortion trade-off 
for some value of $n$ and the quantum efficiency. 
\par
\begin{figure}[h]
\includegraphics[width=0.49\textwidth]{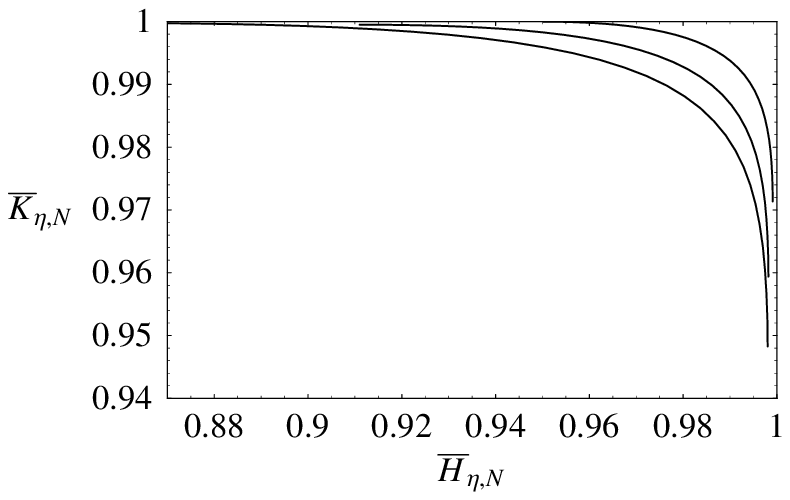}
\includegraphics[width=0.49\textwidth]{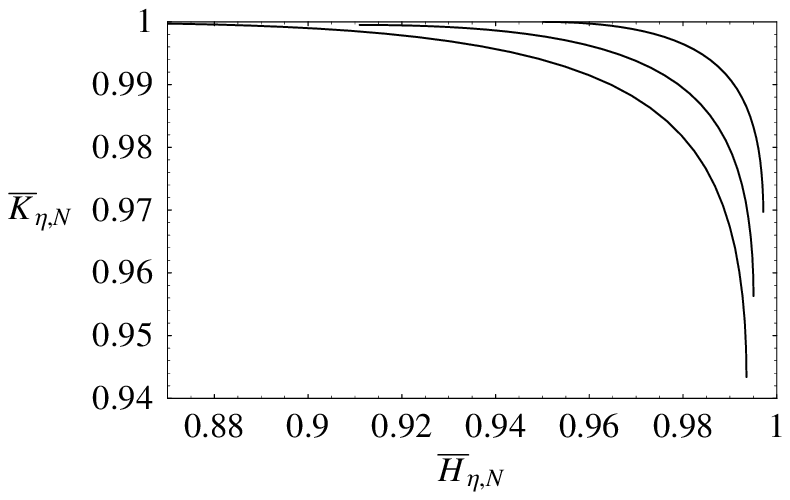}
\caption{Optimized estimation/distortion trade-off for 
thermal sets of Fock number states. (Left): $\eta=0.9$. 
(Right): $\eta=0.8$. In both plots we show the trade-off 
for (from right to left) $N=0.5, 1, 2$. .\label{f:nth}}
\end{figure}

If the different number states are sent to the input according to a 
thermal probability
distribution 
\begin{equation}
p_n = \frac{1}{1+N} \left( \frac{N}{1+N} \right)^n\,, \quad n \ge 0\,.
\end{equation}
the average fidelities are given by 
\begin{eqnarray}
\!\!\!\!\!\!\!\!
\!\!\!\!\!\!\!\!
\overline{H}_{\eta}(\phi) = \sum_{n=0}^{\infty} p_n
H_{\eta,\kappa}^{(n)}(\phi) \qquad 
\overline{K}_{\eta,g}(\phi) = \sum_{n=0}^{\infty} p_n
K_{\eta}^{(n)}(\phi)\:.
\label{hkn}
\end{eqnarray}
We have not been able to find a closed analytical 
form for the corresponding trade-off. 
In Fig. \ref{f:nth}  we show the trade-off, as obtained by numerical
evaluation of Eqs. (\ref{hkn}), upon a suitable truncation of the 
Hilbert space, for different values of $N$ and $\eta$.
%%%%%%%%%%%%%%%%%%%%
\section{Conclusions}\label{s:outro}
We have analyzed a linear optical scheme for the extraction of 
information about a set of coherent states with minimum disturbance.
We have shown that non universal protocols improve the trade-off
compared to universal one. 
We have also introduced the estimation/distortion trade-off to 
assess the indirect measurement of the $Q$-function of a generic
set of states and explicitly evaluated it for coherent and Fock
number states. 
%%%%%%%%%%%%%%%%%%%%
\ack
This work has been supported by MIUR through the project
PRIN-2005024254-002. We thank the referees for useful suggestions.
%%%%%%%%%%%%%%%%%%%%

%%%%%%%%%%%%%%%%%%%%%%%%%%%%%%%%%%%%%%%%%%%%%%%%%%%%%%%%%%%%%%%%%%%%%%
\end{document}